\documentstyle[aps,prl,epsfig,twocolumn]{revtex}

\def\be{\begin{equation}}
\def\ee{\end{equation}}
\def\bea{\begin{eqnarray}}
\def\eea{\end{eqnarray}}
\def\bma{\begin{mathletters}}
\def\ema{\end{mathletters}}
\def\bg{\begin{guess}}
\def\eg{\end{guess}}
\def\C{\hbox{$\mit /$\kern-.6em$\mit C$}}
\def\one{\hbox{$\mit I$\kern-.6em$\mit I$}}

\newcommand{\eins}{\mbox{$1 \hspace{-1.0mm}  {\bf l}$}}
\newcommand{\half}{\mbox{$\textstyle \frac{1}{2}$}}
\newcommand{\shalf}{\mbox{$\textstyle \frac{1}{\sqrt{2}}$}}
\newcommand{\ket}[1]{ | \, #1  \rangle}
\newcommand{\bra}[1]{ \langle #1 \,  |}

\newcommand{\scal}[2]{\bra{#1}#2\rangle}
\newcommand{\abs}[1]{ | \, #1 \,  |}

\newcommand{\calo}{\mbox{${\cal O}$}}
\newcommand{\rhota}{\mbox{$\rho^{T_A}$}}
\newcommand{\bracket}[2]{\mbox{$\langle {{#1}} \mathrel{ | {\vphantom
        {{#1} {#2}}} \kern-\nulldelimiterspace} {{#2}} \rangle$}}
\newcommand{\rem}[1]{}

\def\R{\hbox{$\mit I$\kern-.277em$\mit R$}}
\def\N{\hbox{$\mit I$\kern-.277em$\mit N$}}
\def\C{\hbox{$\mit I$\kern-.7em$\mit C$}}
\def\un{\leavevmode\hbox{\normalsize1\kern-4.6pt\large1}}

\begin{document}
\draft

\title{Separability and distillability 
in composite quantum systems -- a primer --}
\author{M. Lewenstein$^1$, D. Bru\ss$^1$, J.I. Cirac$^2$, B. Kraus$^2$, M. Ku\'s$^3$, J. Samsonowicz$^4$, A. Sanpera$^1$ and  R. Tarrach$^5$}

\address{$^1$Institut f\"ur Theoretische Physik, Universit\"at Hannover,
D-30167 Hannover, Germany\\
$^2$ Institut \"ur Theoretische Physik, Universit\"at Innsbruck, A--6020
Innsbruck, Austria\\
$^3$ Centrum Fizyki Teoretycznej, Polska Akademia Nauk, 02-668 Warsaw,
Poland\\
$^4$ Department of Mathematics, Warsaw
Technical University,00-661 Warsaw, Poland\\
 $^5$ Departament d'Estructura i Constituients de la Materia, 
Universitat Barcelona, Spain }

\date{\today}

\maketitle

\begin{abstract}

Quantum mechanics is already 100 years old, but remains 
alive and full of challenging open problems. On one hand,   
the problems encountered at the frontiers of modern theoretical physics 
like Quantum Gravity, String Theories, etc. concern Quantum Theory, 
and are at the same time 
related to open problems of modern mathematics. 
But even within  non-relativistic quantum 
mechanics itself there are fundamental unresolved problems 
that can be formulated in elementary terms. These 
problems are also related to challenging open questions 
of modern mathematics; linear algebra and 
functional analysis in particular. 
Two of these problems will be discussed in this article: 
a) the separability problem, i.e. the 
question when the state of a composite quantum system does not 
contain any quantum correlations or entanglement and  b) 
the distillability problem, i.e. the question when the  state of 
 a composite quantum system can be transformed  to
an entangled pure state using local operations (local refers here to 
component subsystems of a given system). 

Although many results concerning the above mentioned problems 
have been obtained (in particular in the last few years in the framework
of Quantum Information Theory), both problems remain until 
now essentially open.
We will present  a primer on the current state of 
knowledge concerning these problems,
and discuss the relation of these problems to 
one of the most challenging questions of linear algebra:
the classification and characterization  of positive operator maps.  
\end{abstract}


\normalsize

\section{ Introduction}

Quantum Mechanics celebrates in this year its first century
of life. In October 1900, Max Planck presented to
 the ``Deutsche
Physikalische Gesellschaft''
his seminal papers: ``\"Uber eine Verbesserung der Wienschen
Spektralgleichung'' and   ``Zur Theorie des Gesetzes der 
Ener\-gie\-ver\-tei\-lung
im Normalspektrum''\cite{planck}.

While there are no doubts about the success of Quantum Mechanics in
explaining  - beautifully - many of the problems concerning a variety of
physical topics,  it is worth stressing that  Quantum Theory is
by no means {\it pass\'ee}. It is nowadays still an open theory with
several  challenges which compel both Physics and Mathematics. Just a
few years ago, it was common belief that the``really big" unsolved
problems of theoretical physics pertained only to the domain of Quantum
Gravity, that is the   conjunction of Quantum Field  Theory and General
Relativity.  Quantum Gravity and  String Theory are
closely related to unsolved challenges of modern  mathematics, in
particular  concerning  algebraic topology and algebraic geometry.
However, to the surprise of many, the emerging field of Quantum
Information Theory\cite{reviews} has shown that even in the
``simple" non-relativistic Quantum Theory there exist still fundamental 
open problems. The characterization of entanglement, and more
specifically, the characterization of separability and distillability of quantum states are
among these. Again, they are directly linked to unsolved
challenges  of mathematics concerning  linear
algebra and geometry, functional analysis and, in particular, the theory of
C$^{\star}$-algebras\cite{cstar}.

In this paper we present some of these new open problems and report on
the recent progress concerning them. The paper does not intend to be
a review article on the subject of quantum entanglement, 
but rather an introduction -- a primer -- on the subject\footnote{This paper has been presented by M.
Lewenstein at the conference ``Quantum Optics", K\"uhtai, January 2000.}. 
It contains nevertheless some new
results: we present here two novel  separability checks, and
some new results concerning distillability of density matrices that possess a 
non-positive partial transpose. 
The presentation of these  new results and their proofs
require the introduction  of some  technical formalism.
Therefore, they have been included as  Appendices.\\
  
The paper is organized as follows: First, in Section II
we explain  what  the entanglement problem means. In Section III we discuss 
the problem of
separability, that is how to define and discriminate those states 
that contain only classical correlations and no quantum correlations.
In Section IV we focus on the problem of distillability, that is, 
the possibility--
by using local operations and classical communication only--  
to ``distill" from a given ensemble of copies of a given mixed state a
maximally entangled pure state. 
In that section we also present the current state-of-the-art. 
Finally, our  summary remarks 
are contained  in Section V.

\section{The entanglement problem}

In order to explain what  the entanglement problem means we
 first have to specify that in the
following we will consider composite quantum systems\cite{Peres}. 
Physical states of such systems  are in
general mixed and can be represented by density matrices, i.e. hermitian,
 positive definite linear operators
of trace one, acting in the Hilbert space  $\cal H=\cal {H}_A \otimes \cal {H}_B\otimes\ldots$, which is a
tensor product of Hilbert spaces corresponding to subsystems $A,B,\ldots$ 
of the considered system.

Given a quantum state
$\rho\in\cal H$, an apparently innocent question as {\it {does this state contain quantum 
correlations?}} will  in general be very hard 
(if not  impossible!) to answer.
First of all, what does it mean that a quantum state does or does not contain quantum correlations?
The answer seems to be 
straightforward: a system contains quantum correlations if
the observables of the different subsystems are correlated, and 
their correlations cannot be reproduced by any means classically. 
That implies that some form of
{\it non-locality} \cite{Peres} is necessary  in order to account for such  
correlations.  
For pure states
-- described by projections on a single vector acting on the Hilbert 
space ${\cal H}$ of the composite
system -- it is relatively easy to check if the correlations 
that they contain are classical, or not. 
For instance, it is enough to check 
if some kind of Bell inequality\cite{bell} is violated to assert 
that the state
contains quantum correlations. In fact, there are many 
different ``entanglement''- criteria and all of them  reveal 
equivalent forms  of the non-local  character of the entangled pure states. 
For example, 
the demonstration that no local hidden variable (LHV) can account 
for the correlations between the observables in each subsystem 
is an equivalent definition of non-locality\cite{werner}. 

We know nowadays that these equivalences may fade away when one deals 
with mixed states. Contrary to a pure state, a mixed state
can be prepared in many different ways. The fact that we cannot 
trace back how it was prepared prevents us from extracting all 
the information contained in the state. 
As a consequence, we lack (nowadays)  
general   ``entanglement'' criteria that allow us to check if 
the correlations present in the system are genuinely quantum, or 
not. Despite the fact that many entanglement measures have 
been introduced, we do not
know a ``canonical'' way of quantifying the entanglement\cite{entaquan}.
Furthermore,  
different manifestations of  non-locality are known to be not
equivalent. For instance, Werner\cite{werner} introduced a family of mixed states that 
do not violate Bell-type inequalities 
(they admit a local hidden variable model), but nevertheless are non-local. The question whether 
there exists a violation of Bell
inequalities in the, so-called, strong sense (where the observables take 
``unphysical" values)
for all PPT entangled
states (which we define below) remains open\cite{terh99}.

Therefore, the entanglement problem can be outlined as: 
{\it What does it mean that a given (mixed) state $\rho$ contains 
or does not contain quantum
correlations?}

\section{The separability problem}

An essential  step forward to understand what does entanglement mean 
is to  discriminate first the states that contain 
classical correlations only (or no correlations at all). 
These states are termed separable states, 
and their mathematical characterization has been formulated by
Werner\cite{werner}. We shall restrict ourselves here 
to the most simple composite systems: bipartite systems 
(with two subsystems traditionally denoted as Alice and Bob) of finite, but 
otherwise arbitrary dimensions. 
The states of bipartite systems are described by positive definite hermitian 
density matrices (with normalized trace) $\rho$, i.e.  $\rho\ge 0$, $\rho=\rho^{\dagger}$ and
${\rm Tr}\rho=1$.   The density matrices  act on the Hilbert space of the composite system 
$\cal H=\cal {H}_A \otimes \cal {H}_B$. Without loosing generality we will assume that
dim ${\cal {H}}_A=M\ge 2$ and dim ${\cal {H}}_B=N \ge M$.

The most simple examples  of  separable states are just  product states, i.e. 
$\rho=\rho^A\otimes\rho^B$ ($\rho^A$ acts on  ${\cal H}_A$, and 
$\rho^B$ acts on ${\cal H}_B$).
These states contain no correlations whatsoever. A straightforward  extension of
product states are the states that contain only classical correlations.  
Werner\cite{werner} provided  us with the following  operational definition of separability:

\newtheorem{guess}{Definition}
\bg 
A given state $\rho$ is separable if and only if
\be
\rho=\sum_{i=1}^{k}p_i \rho_i^{A}\otimes \rho_i^B\ ,
\label{werner}
\ee
where $\sum_ip_i=1$, and $p_i\ge 0$.
\eg 
The above expression means  that $\rho$ can be written as a convex combination of product states.
Equation (1) has a clear physical meaning. The state $\rho$ can be prepared
by Alice and Bob by  means of local operations (unitary
operations, measurements, etc.) and classical communication (LOCC).
If $\rho$ is separable the system does not contain quantum correlations. 
In spite of the definition, the characterization of such states is
a rather arduous task. This is so among other facts because, in general, 
even for a given  separable generic matrix we do not have 
an algorithm to decompose it according to Eq. (\ref{werner}). 
Thus, the separability
problem, perhaps even more basic and fundamental than the entanglement problem
can be formulated  as:
{\it {Given a composite quantum state described by $\rho$, 
is it separable or not?}}

Before proceeding further we introduce here the definitions that we shall use
throughout the paper. Given a density matrix $\rho$,  we denote by $K(\rho)$,
$R(\rho)$, and $r(\rho)$ the kernel, the range and the rank of the matrix $\rho$ defined as:

\begin{guess}
Kernel $K(\rho)=\{\ket\phi: \rho\ket\phi=0\}.$
\end{guess}

\begin{guess}
 Range  $R(\rho)=\{\ket\phi:\exists \ket{\psi}:
\ket \phi=\rho\ket\psi$\}.
\end{guess}

\bg
Rank $r(\rho)={\rm dim}\; R(\rho)=NM-{\rm dim}\, K(\rho).$
\eg

\noindent Let us introduce also the operation of ``partial transposition'' 
that will be used throughout the paper and is defined as:\\

\bg
The partial transpose of $\rho$ means the 
transpose only with respect to one of the subsystems. If we express $\rho$ in
Alice's  and Bob's orthonormal product basis:
\bea
\rho&=&\sum_{i,j}^{M}\sum_{k,l}^{N}\bra{i,k} \rho\ket{j,l}\ket{i,k}\bra{j,l}\nonumber\\
&=&\sum_{i,j}^{M}\sum_{k,l}^{N}\bra{i,k} \rho\ket{j,l}{\ket{i}}_A\bra{j}\otimes
{\ket{k}}_B\bra{l},
\eea
then the partial transposition with respect to Alice is expressed as:
\be
\rho^{T_A}=\sum_{i,j}^{M}\sum_{k,l}^{N}\bra{i,k} \rho\ket{j,l}\ket{j}_A
\bra{i}\otimes\ket{k}_B\bra{l}.
\ee
\eg
Note that $\rhota$ is basis-dependent, but its spectrum  is not.  
The partial transpose $\rhota$ may be 
$\geq 0$, but does not have to be!
As
$({\rhota})^{T_B}=\rho^{T}$, and as $\rho^{T}\geq 0$ always 
holds, positivity of $\rhota$ implies positivity of 
$\rho^{T_B}$ and vice versa.

A major step  in the characterization of the separable states was done by 
Peres\cite{peres1} and the Horodecki family\cite{horo}. Peres provided a 
``userfriendly" and very powerful necessary condition for separability. Later on, Horodecki's demonstrated that this
condition is also sufficient for composite Hilbert spaces of dimension
$2\times 2$ and $2\times 3$. Their results are enclosed in the following two
theorems:

\newtheorem{guess1}{Theorem} 
\begin{guess1}
If $\rho$ is separable then  $\rhota\geq 0$.
\end{guess1}
A matrix that verifies the above theorem is  termed ``PPT'' 
for positive partial transpose. Notice that being a PPT state is a 
necessary condition for separability.

\begin{guess1}
If $\rhota\geq 0$ in spaces of dimensions $2\times 2$ or
$2\times 3$ then $\rho$ is separable.
\end{guess1}
In general, there exist PPT states $\rho$ (i.e. states with $\rhota\geq 0$) 
which are not separable in $M\times N$ spaces
($M=2, N\geq 4$ or $M\geq 3$) \cite{pawel}. 
The PPT entangled states have been  termed ``bound entangled
states'' to distinguish them from the ``free entangled states''. 
This latter names are associated 
with the distillability property, which we will discuss in the later 
sections of this paper. ``Bound entangled states" are entangled, however, no
matter how many copies of them we have, these states cannot be ``distilled" via
local operations and classical communication to the form of a pure entangled state\cite{bound}. 
We encounter thus new problems such as:
 {\it How can one distinguish 
a separable state $\rho$ from a PPT state $\rho$? Are all non-PPT states 
(NPPT states) ``free entangled" i.e.
distillable?} 
But before trying to answer these questions (that will bring us directly
to the problem of distillability), we will first present a physical explanation
of what separability means,  and discuss shortly the recent progress
 concerning the quest for separability
criteria.

\subsection{Physical interpretation of separability}

Let us now interpret the condition of positive partial transposition 
from a physical point of view. 
We start by considering symmetry transformations in the Hilbert space of each subsystem.
Wigner's theorem\cite{wigner} tells us that every symmetry transformation
is necessarily implemented by a unitary ($U$) or anti-unitary ($A$) matrix. 
The tensor product of  a unitary and an anti-unitary transformation 
$U_A\otimes A_B$ (or $A_A\otimes U_B$) 
results in  a transformation which is neither unitary, nor anti-unitary
in $\cal H=\cal H_A\otimes\cal H_B$, and  whose action on 
a general ket of the composite system 
$\ket{\psi}\in \cal H$, furthermore, cannot be properly defined. 
However, its action on a product ket $\ket{e,f}\equiv\ket{e}\otimes\ket{f}$, 
(where $\ket{e}\in {\cal H}_A$ and $\ket{f}\in {\cal H}_B$) is, 
apart from a phase
 ambiguity, well defined. Thus, the action of a combined transformation of 
the type $U_A\otimes A_B$ on projectors corresponding to pure product states 
is well defined without any ambiguity. As a separable state $\rho_s$ can always
 be rewritten as a statistical mixture of product vectors (see Def. 1) it is clear
 that under the combined transformation $U_A\otimes A_B$ 
(or $ A_A\otimes U_B$), $\rho_s$ transforms into:
\be
\rho_s \rightarrow \rho_s^{'}=\sum_{i=1}^{k}\,p_i \left (\ket{e_i^{'}}
\bra{e_i^{'}}\otimes \ket{f_i^{'}}\bra{f_i^{'}}\right )
\ee
where $\ket{e_i^{'}}\equiv U_A\ket{e_i}\in {\cal H}_A$ ; $\ket{f_i^{'}}\equiv 
A_B\ket{f_i}\in {\cal H}_B$. Therefore, $\rho_s^{'}$ describes also a physical state 
so that $\rho_s^{'}$ is a positive definite hermitian matrix 
(with normalized trace). 
This is what characterizes separable states: that any local symmetry
transformation, which obviously transforms local (in this context local refers
to each of the subsystems) physical states into local 
physical states, also transforms the composite global state into another 
physical state.
 
There exists only one independent anti-unitary symmetry\cite{wigner}, and its physical meaning is well known:
time reversal. Any other anti-unitary transformation can be expressed in terms of time reversal (as the
product of a unitary matrix times time reversal). Thus separability 
of composite systems implies  the lack of correlation between 
the time arrows of  their subsystem. In other words: 
given a  separable composite state, reversing time in one of its subsystems 
leads again to a physical state \cite{sanpera97,bush97}. 

\subsection{Quest for separability criteria and checks}

In the recent years there has been a growing effort 
in searching for necessary and 
sufficient separability criteria and checks. 
Several  necessary conditions for separability are known: 
Werner has derived a condition based on the analysis of 
local hidden variables (LHV) models and the mean 
value of the, so-called,
flipping operator
\cite{werner}, the Horodecki's have proposed a necessary criterion based 
on the so-called $\alpha$-entropy inequalities\cite{alfa}, etc... 
Quite recently, a general and sufficient condition for separability
was discovered by the Horodecki family in terms of positive maps. A map is defined positive if it maps positive operators into positive operators. 
The condition
found by the Horodecki's, states that $\rho$ is separable iff the 
tensor product of any positive map acting on one subsystem {\it A} and the 
identity acting on other subsystem {\it B} maps $\rho$ into
a non-negative operator. 
This definition, however, involves the characterization of the set
of all positive maps which is {\it per se} a major task. Later on  
the reduction criterion of separability  was introduced
\cite{xor,Adami}: 
\newtheorem{crit}{Criterium} 
\begin{crit}
If $\rho$ is separable then
the map $(\Gamma_A\otimes\eins):
\rho\rightarrow (\eins_A\otimes {\rm Tr_A} \rho)-\rho$ must be positive.
\end{crit}
Violation of this criterion is sufficient for entanglement to be free.  
Following the reduction criterion, a simple and still quite powerful sufficient condition
for distillability
was provided in Ref.\cite{rank2}, where P. Horodecki {\it et al.}
showed that  if the rank of at least one of 
the reduced density matrices $\rho_A={\rm Tr}_B\rho$
and  $\rho_B={\rm Tr}_A\rho$ exceeds
the rank of $\rho$, then $\rho$ 
is distillable, {\it ergo} is non-separable and NPPT.
In particular, it was concluded in Ref. \cite{rank2}
that there is no bound entanglement of rank 2. 

Sufficient conditions for separability are also known. In 
Ref. \cite{karol} it was proven that any
state close enough to the completely random state $\pi=I/NM$ is separable. 
In  \cite{karol} there were also given  the first
quantitative bounds for the radius of the ball surrounding $\pi$ that does not contain any entangled state.
Much better bounds were found in the following works\cite{guifre}, where it was proven that a full
rank mixed state is separable 
 provided that its smallest eigenvalue is greater or equal to $(2+MN)^{-1}$.
 
In Ref. \cite{pawel}, in which the first explicit examples 
of entangled states
with PPT property were provided, another necessary 
criterion  of separability was formulated.
According to this criterion:
\begin{crit}
If the state $\rho$ acting on a finite dimensional Hilbert space is separable 
then there  must exist a set of product vectors $\{|e_i,f_i\rangle\}$ that spans the range $R(\rho)$ such
that the set of partially complex conjugated product states $\{|e^*_i,f_i\rangle\}$ spans the range of $\rhota$.
\end{crit}
The
analysis of the range of the density matrices, initiated by P. Horodecki,  turned out to be  very fruitful,
leading, in particular, to  the algorithm of optimal decomposition of mixed states
into the separable and inseparable part \cite{M&A,sanpera}, and
to  systematic methods of constructing  
examples of PPT entangled states 
with no product vectors in their range,
using either  so-called unextendible product bases (UPB's) \cite{UPB,terhal},
or the method described in \cite{dbap}.

In the Appendices A and B we present two novel separability criteria and checks. 
One provides a necessary
condition for separability, or rather a sufficient condition for entanglement. It detects the 
non-separability of the UPB states. 
The other criterion, or rather separability check, detects all separable
states that are the convex combinations of two product states.

\subsection {Recent progress in the separability problem}

Despite many efforts and seminal results obtained in the recent years, 
the problem of separability remains essentially open.  
Recently, a considerable progress of in the study of PPT entangled states 
has been made \cite{2xn,mxn}. The results obtained allow us to hope 
to develop a systematic way of 
constructing optimal criteria for separability in arbitrary Hilbert spaces\cite{witnes}.

Our method  employs the idea of ``subtracting projectors on product vectors" 
\cite{M&A,sanpera}: if there
exists a product vector $|e,f\rangle\in R(\rho)$ such that $|e^*,f\rangle\in R(\rhota)$, the
projector onto this vector (multiplied by some $\lambda>0$) can be subtracted 
from $\rho$, such that the
remainder is positive definite and PPT.   Our results   can be divided into three groups. 

First, we have
studied and found separability criteria  for density matrices of sufficiently low dimensional rank.  Also constructive  algorithms to decompose optimally (with the
smallest possible  number of terms)  the separable matrix according to Eq. (1) - i.e. in product states -
have been provided for low rank matrices.    For the general case of composite systems with the Hilbert
space
$\cal{H}^M\otimes\cal{H}^{N}$ ($M\le N$) our findings are essentially contained  in the following two
theorems:
\begin{guess1}
If $\rho$ is PPT such that $r(\rho)=k\le N$ and $\rho$ cannot be
embedded in a $2\times (k-1)$ dimensional space then $\rho$ is separable.
\end{guess1}
In particular when $\rho$ has rank $N$ there exist typically exactly $N$ product vectors $|e_i,f_i\rangle$
in the range of $\rho$ such that $|e^*_i,f_i\rangle\in R(\rhota)$; 
$\rho$ is then a convex combination of
projections onto these vectors.
\begin{guess1}
If $r(\rho) + r(\rhota)\leq 2NM-N-M+2$ then typically there exists a 
{\bf{finite}} number of product vectors $\ket{e,f}\in R(\rho)$ such that
$\ket{e^{\star},f }\in R(\rhota)$.
\end{guess1}

These product vectors are the only possible candidates to 
appear in the decomposition of Eq. (1).  Finding them requires
solving a system of polynomial equations. After these equations are solved, one can check whether $\rho$
has the decomposition (1). The problem is infinitely  easier than the original one since
we know now all possible projectors that can be used, and we know that their number  is finite. 
In fact, checking in such a situation whether $\rho$  is separable or not can be done   
in  a finite number of  computational steps!

Second, we have studied the structure and generic form of low rank PPT entangled matrices. To study low rank PPT entangled matrices has a twofold purpose. On one hand, the complexity
of the problem is reduced, and therefore it is possible to find separability criteria. 
But, perhaps the most important is the fact that, given a 
density matrix $\rho$, one can always decompose it as :

\bea
\rho&=& \Lambda \rho_{sep}+(1-\Lambda)\delta \rho\ ,\\
\rhota&=& \Lambda \rhota_{sep}+(1-\Lambda)\delta \rhota\ ,
\eea
where $\rho_{sep}=\sum_i\Lambda_i P_i$ is a separable state, $P_i$ are 
projectors onto product states
 and  $\Lambda$ is maximal. All the information concerning entanglement 
is then contained in the
remainder $\delta \rho$ ($\delta \rhota$) which has low rank and can be 
termed as a ``pure" 
PPT entangled state, or ``edge" PPT entangled state. 
This state has a property that no projection onto the product state
can be subtracted from it, keeping the rest positive definite and PPT. 
Formally, there exist no product vectors $|e,f\rangle\in R(\delta\rho)$ such 
that $|e^*,f\rangle\in R(\delta\rhota)$. The ``edge" states violate in the
extremal sense the Criterion of Ref. \cite{pawel}.    
The problem of the separability reduces now to the
problem of separability of the ``edge" states, 
and to the question whether a given mixture of an ``edge" and 
a separable state is separable or not.    

In other words, any matrix $\rho$ can be decomposed in a separable part
(that contains product vectors in the range) and a remainder, 
which is the ``edge" state.   $\delta\rho$ and $\delta \rhota$ are  low rank matrices that 
contain all the information related to entanglement. 
Obviously,  knowing the structure of those matrices   
is therefore of capital importance.

Finally, let us mention a different approach to the entanglement
problem, that is based on the so-called entanglement witnesses. An entanglement witness is
an observable $E$ that reveals the entanglement of an entangled  density matrix
$\rho$.   B. Terhal\cite{terh99,terhal} introduced  entanglement witnesses through the following
theorem:

\begin{guess1} 
If $\rho$ is entangled then there exists 
an {\it {entanglement witness E}} such that 
\bea
&{\rm Tr}&(E\rho_{sep})\geq 0,\,\\
&{\rm Tr}&(E\rho)<0
\eea
for all separable matrices $\rho_{sep}$.
\end{guess1}
Entanglement witnesses represent -- in some sense -- a kind of Bell inequality 
which is violated by the entangled
state $\rho$. Each entanglement witness $E$ on an $M\times N$ space defines a positive map ${\cal E}$
that transforms positive operators on an $M$ or $N$--dimensional Hilbert space 
into positive operators on
an $M$ or $N$--dimensional space\cite{jamio}. The maps corresponding to  entanglement witnesses are positive, but
not completely positive, and in particular their extension to $M\times N$ spaces 
allows to ``detect" the
entanglement of $\rho$. The maps corresponding to entanglement witnesses for PPT states are, moreover, 
non decomposable: they cannot be represented as a combination of completely positive maps 
and partial transposition.

For every ``edge" state it is possible to construct an entanglement 
witness\cite{witnes} as: 
\be
E=P_{K({\rho})}+ (P_{K({\rhota})})^{T_A}- \epsilon ,
\ee
where $P_{K({\rho})}$,  $P_{K({\rhota})}$ are projections onto the kernel of $\rho$ and
the  kernel of $\rhota$, 
and $0<\epsilon={\rm min}\langle e,f |P_{K(\rho)}+ (P_{K(\rhota)})^{T_A}|e,f\rangle$, where 
the minimum is taken over all possible product vectors.

We have not only been able to find entanglement witnesses, and the corresponding non--decomposable 
positive maps for arbitrary ``pure" or ``edge" PPT states, but also to optimize them in a certain sense
 \cite{witnes}.
Optimized entanglement witnesses detect significantly more entangled PPT states than the non--optimized ones.

We hope very much that these studies will allow us to characterize  extremal points in the convex set of
PPT entangled matrices, and then to characterize 
the  extremal points in the convex set of positive maps~\cite{lewenwit}.
If this program is realized, 
 the separability problem will be solved. So far, however, only  the first steps
have been done and the problem remains open and challenging.

\section {The Distillability Problem}
\label{dist}
On having said that, we shall attack now the related problem of the  distillability of 
mixed quantum states. 
For many applications in quantum information processing \cite{ekert}
and communication  one needs a
maximally entangled state, that is, 
a state which in $M\times N$ dimensional space can be brought  by a local change
of basis to the form
\be
\ket{\Psi_{max}}=\frac{1}{\sqrt{M}}\sum_{i=1}^{M}\ket{i,i},
\ee
 which is shared between two parties. 

Although in principle one can create pure and maximally entangled states, in  realistic situations
any pure state  will evolve to  a mixed state
due to its interaction with the environment. A standard example concerns a 
situation when two entangled particles (photons, atoms,...)  representing the two subsystems are sent
from the source to the two involved  parties, Alice and Bob, through noisy channels.
In order to overcome the noise created during the transmission, the idea of 
distillation and purification, i.e. enhancement of the given non-maximal mixed  entanglement 
by local operations and classical communication (LOCC) was proposed by Bennett {\it et al.}
\cite{bennet}, Deutsch {\it el al}~\cite{deutsch} and Gisin ~\cite{gisin}. Again,
for Hilbert spaces of composite systems of dimension lower or equal to 6, any
mixed entangled state can always be distilled to its pure form. Since for such systems entanglement
is equivalent to the NPPT property, we conclude that for $2\times 2$, and for $2\times 3$ systems,
all NPPT states are distillable\cite{lowdis}. However, it was shown by 
Horodecki family
\cite{bound} that in higher dimensions there exist
 states (namely PPT entangled states), termed as bound
entangled states, which cannot be distilled, in contraposition 
to free entangled states\cite{aktyw}. The distillability  problem can be formulated as: 
{\it{ Given a density matrix
$\rho$, is it or is it not distillable?}} 

Let us now define
the distillability property, first on an intuitive, and then 
 on
a more formal basis.

\begin{guess}
$\rho$ is distillabe if by performing LOCC on 
some number $K$ of copies $\rho$, Alice and Bob can distill a state arbitrary
 close to 
$\ket {\Psi_{max}}$, i.e.
\be\rho\otimes....\otimes\rho\longrightarrow  \ket{\Psi_{max}}\
\bra{\Psi_{max}}.
\ee
\eg

The above definition is not very precise -- it requires to specify what the LOCC can do with $K$ 
copies of $\rho$, and does not  give any practical advice about how to answer the question
of distillability. Fortunately we can use the theorem of Ref.\cite{bound}, 
which states that instead of studying the whole set of possible LOCC, 
in order to determine the distillability of a given density matrix it is
sufficient to study projections on a $2\times2$ dimensional subspace of the Hilbert space in which $\rho$
acts. The theorem is very useful since it reduces the problem of distillability to a very precisely stated
mathematical question; in fact from  now on we will use it as a definition of distillability.

\begin{guess1}
$\rho$ is distillable iff there exists
a number of copies  $K$, and
a projector $P_{2\times 2}$ onto a $2\times 2$-dimensional 
 space spanned by:
\bea 
&\ket{e_i}&\in \underbrace {\cal{H}_{A}\otimes...\otimes \cal{H}_{A}}_
{K-times}\;, \ i=1,2\,,\\
&\ket{f_i}&\in \underbrace {\cal{H}_{B}\otimes...\otimes \cal{H}_{B}}_
{K-times}\;,\  i=1,2\,,
\eea
such that the projection
\be
\sigma=P_{2\times 2}\rho^{\otimes K}P_{2\times 2},
\ee
is NPPT (i.e. is distillable).
\end{guess1}
An alternative way of formulating the above theorem  is the following: 
$\rho$ is distillable iff there exists a state $\ket{\psi}$ from a  
$2\times 2$-dimensional subspace,
\be
\ket{\psi}=a\ket{e^{\star}_1}\ket{f_1}+b\ket{e^{\star}_2}\ket{f_2}\ ,
\label{psi2by2}
\ee
such that $\bra{\psi}(\rho^{T_A})^{\otimes K}\ket{\psi}<0$ for some $K$.

The idea of the proof of the above theorem  is the following: if $\rho$ is distillable it means
that one can produce a maximally entangled state, and it is then easy to project (using local
projections) that state  onto a pure state in a $2\times 2$-dimensional subspace. 
On the other hand, if 
there is a $\ket{\psi}$ as in equation (\ref{psi2by2}), such that 
$\bra{\psi}\rho^{T_A}\ket{\psi}<0$, then one can first project onto the $2\times 2$ subspace to which 
$\ket{\psi}$ belongs. 
This is a $2\times 2$ subspace in  which the projected matrix is NPPT, {\it ergo} it 
is distillable. We can then distill several
maximally entangled state in this subspace, rotate them  unitarily and locally, and combine to a
maximally entangled state in the whole space.

Let us now ask what does the criterion of partial transposition,-- which plays
an important role in the separability problem as we have seen before--,
tell us about the distillability problem?

\begin{guess1}
If $\rhota\ge 0$ ($\rho$ is PPT) then $\rho$ is
not distillable \cite{bound}.
\end{guess1}

\begin{guess1} 
 If $\rhota\not\ge 0$ ($\rho$ is NPPT) in 
dimensions
$2\times 2$, $ 2\times 3$ then $\rho$ is distillable\cite{lowdis}.
\end{guess1}
The later holds also  for $2\times N $ systems, see  
\cite{duer}.

\subsection{Recent progress in the distillability problem}

At the end of the last section we have seen that every density
matrix with a positive
partial transpose cannot be distilled, and that for  low dimensions
the converse is true. In this section we want to discuss the
conjecture that in higher dimensions there are states with a
non-positive partial transpose, which are however non-distillable
\cite{duer,divi}.
In other words, non-positivity of the partial transpose seems to be
a {\it necessary}, but {\it not} a 
{\it sufficient} condition for distillability.

The states which are believed to be non-distillable\cite{duer}, belong to 
a one-parameter family, which lives in dimension $N\times N$:
\be
\rho(\alpha) =\frac{1}{m(\alpha)} (P_S + \alpha P_A ) 
\ \rm{with} \ \alpha\geq 0\ ,
\label{can}
\ee
where $P_S$ and  $ P_A$ denote projectors onto the symmetric
and antisymmetric subspace, respectively, and $m$ is some
normalization. This family is generic in the sense that {\it every}
density matrix in $N\times N$ can be depolarised locally to a
state from our family. In this sense these states are nothing more but Werner states\cite{werner}
defined for $N\times N$ systems.
\par 
The partial transpose of $\rho(\alpha)$ is given by
\be
\rhota(\alpha) =\frac{1}{n(\beta)} (1 - \beta P) \ ,
\ee
where $n(\beta)$ is the normalization, $P$ is the projector onto a maximally entangled state
$|\Psi_{max}\rangle=\sum_{i=1}^N|ii\rangle/\sqrt{N}$, 
and the relation between $\alpha$ and $\beta$ is:
\be
\beta = \frac{N(\alpha -1)}{\alpha +1} \ \ \rm{with} \ -N\leq\beta\leq N\ .
\ee
Note that  for $\beta\le 1$ the matrix $\rhota$ is positive definite , i.e. $\rho$ is PPT and thus is not
 distillable. For $\beta>1$ one finds that
$\rhota$ is not positive definite,  i.e. $\rho$ is NPPT and the question of distillability is open.

The nice thing about the considered family of states is that
if we show that:\\
\noindent (i) $\rho$ is distillable for all 
$\beta > 1$, then all $\rho$
with NPPT are distillable, because all $\rho$ can be reduced to the ``canonical" form (\ref{can}). 
\newline
\noindent (ii) if there exists $\beta\leq N$, such that $\rho$ is NPTT and is not 
distillable, then not  all $\rho$ with NPPT are distillable. 
In other words there  exist  
undistillable $\rho$'s with NPPT.
Both alternatives (if proven) would be an extremely important result. At the moment it seems that the second
alternative is true \cite{duer,divi}, but strictly speaking the problem is open.
We shall see below that $\rho$ is distillable for $\beta>3/2$, so that in fact interesting region
of the parameter $\beta$ lies between 1 and 3/2.

Before presenting  some partial results on the way to the complete
proof, which is yet unknown, let us introduce 
 the concept of $K$-distillability, by which 
we define distillability with respect to $K$ copies.

\begin{guess}
$\rho$ is $K$-distillable iff there 
$\exists \ket{\psi}$ as given in Eq. (\ref{psi2by2}) such that
\be
\bra{\psi} (\rhota)^{\otimes K}\ket{\psi}<0.
\label{sub}
\ee
\end{guess}
The basic theorem that we have proven so far determined the region of 
the parameter $\beta$ for
which the matrix $\rho$ of Eq. (\ref{can}) is not $K$-distillable. Our results hold for arbitrary $N\times
N$ systems, but here we specify them to case of two qutrit systems ($3\times3$).

\begin{guess1} 
Let $\rho\sim P_S+\alpha P_A$, i.e. 
$\rhota\sim 1-\beta P$ act in $3\times 3$ space, then:\\
\begin{itemize}
\item $\rho$ is not 1-distillable for $1\leq\beta\leq \frac{3}{2}$;

\item $\rho$ is not 2-distillable for $1\leq\beta\leq \frac{5}{4}$;

\item $\rho$ is K-undistillable for $1\leq\beta\leq \beta_K$, where the best bound for $\beta_K$ obtained so far is:
\be
\beta_K\sim 1+\frac{1}{3^{K/3}K^{1/3}}.
\ee

\end{itemize}
\end{guess1}
As we see, for every $K$ there exist a region of $\beta$ (see Fig.1) in  which $\rho$ is not distillable.
Unfortunately, this region shrinks, however to a point, as $K$ goes to infinity. If we have an arbitrary
number of copies of $\rho$ we cannot say whether we will be able or not to distill it.  
The proof of the above theorem is technical, but essentially simple. Below we sketch the proof concerning 
1-- and 2--distillability

\subsection{1--Distillability}

Let $Q=1-P$ be the projector complementary to $P$, where $P$ is the projector
onto maximally entangled states.
For one copy of the matrix $\rho$ 
we have that for any $\ket{\psi}$ given by Eq. (\ref{psi2by2}):
\bea
\bra{\psi} Q-\beta P \ket{\psi}&=&
\bra{\psi}1-(1+\beta) P\ket{\psi}\ge\nonumber\\
& &1-(1+\beta)\frac{2}{3}\ge 0 
\eea
for  $\beta\le \frac{3}{2}$. The last inequality  follows from the fact that the projection
$|\langle\psi|\Psi_{max}\rangle|^2$ of an arbitrary vector living in a $2\times 2$ subspace onto a maximally entangled
vector in $3\times 3$ space must not be greater than 2/3. 
The above considerations imply also that for $\beta>3/2
$ the matrix is 1--distillable, because there exists a vector in a $2\times 2$ subspace for which
$|\langle\psi|\Psi_{max}\rangle|^2=2/3$. This proves that $\rho$ is distillable for $\beta>3/2$.

\subsection{2--Distillability}

To show that  2 copies of $\rho$ cannot be distilled for some interval of $\beta$, we first observe that
Q is separable, i.e.
\be
Q=\sum_i^Rp_i\ket{e_i,f_i}\bra{e_i,f_i}.
\ee
Denoting by ${\rm Tr}_i$ the trace over the $i$--th copy, and by $Q_i$ -- the projector $Q$ for the $i$--th
copy we get
\be  
{\rm Tr}(\bra{\psi}Q_1\ket{\psi})=\sum_{i=1}^R p_i \langle{\psi}\ket{e_i,f_i}
\langle{e_i,f_i}\ket{\psi}
\ee
where $\ket{\psi}_i=\langle{\psi}\ket{ e_i,f_i}$ is a vector 
in the second copy space with 2 Schmidt
coefficients, i.e. of the form (\ref{psi2by2}). Similar result holds for the second copy. 
 Then, using the results for 1--distillability we
obtain
\bea
\bra{\psi}Q\otimes  (Q-\frac{1}{2}P)\ket{\psi}\ge 0, \\
\bra{\psi}(Q-\frac{1}{2}P)\otimes  Q\ket{\psi}\ge 0.
\eea
 Now, by adding the above results  and dividing by two we obtain 
\be
\bra{\psi}(Q-\frac{1}{4}P)\otimes  (Q-\frac{1}{4}P)\ket{\psi} 
\ge \bra{\psi}\frac{1}{16}P\otimes
P)\ket{\psi}\ge 0,
\ee
so that we see that $\rho$ is not 2--distillable for $\beta\le 5/4$.

\subsection{Distillability in general}

In  the Ref. \cite{duer} we have performed extensive numerical studies and looked for the minimum
of $\bra{\psi} (\rhota(\beta))^{\otimes K}\ket{\psi}$ over all possible $\ket{\psi}$ of the form 
(\ref{psi2by2}). The numerical results indicate clearly that:\newline

\begin{itemize}

\item $\rho$ is not 2-distillable for $1\leq\beta\leq \frac{3}{2}$;

\item $\rho$ is not 3-distillable for $1\leq\beta\leq \frac{3}{2}$.

\end{itemize}
It is a challenging and open problem to understand these results. 
So far we have only achieved 
some progress  in the problem
of 2--distillability. We have proven that the states $\ket{\psi}$ for which $\bra{\psi}
(\rhota(\beta))^{\otimes K}\ket{\psi}=0$, and which (as we know) provide the global minimum of 
$\bra{\psi} (\rhota(\beta))^{\otimes K}\ket{\psi}$ for $\beta\le 5/4$, provide also a local minimum 
of $\bra{\psi} (\rhota(\beta))^{\otimes K}\ket{\psi}$ equal to zero for $3/2\ge \beta>5/4$. The proof of
this fact is presented in the appendix C.
There exists a well-based suspicion that $\rho$ is not distillable in the 
entire region of $1\le \beta\le 3/2$.

\section{Conclusions}

There is only one conclusion of this paper: Quantum Theory is an open
 and challenging area of physics. It 
offers still fundamental and fascinating problems that can be formulated 
at elementary level, and yet they are related to 
challenges of the modern mathematics. 
Particular examples of those are separability and distillability of composite
quantum systems. 

This paper has been supported by SFB 407 and Schwerpunkt 
``Quanteninformationsverarbeitung" of 
Deutsche Forschungsgemeinschaft, by the
ESF PESC Programme on Quantum Information, and by the IST Programme EQUIP.

\begin{figure}[hbt]
\vskip .5truecm
\begin{center}\epsfxsize=0.9 \hsize\leavevmode\epsffile{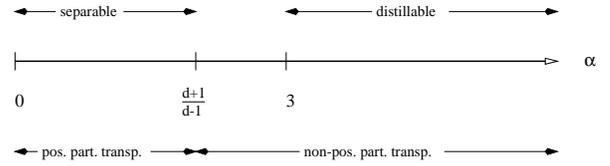}
\end{center}
\vskip .5truecm
\caption[]  
        {\small Separability and distillability properties of $\rho$ 
in arbitrary $d\times d$ dimensional space. In the region 
between separable and distillable $\rho$ is NPPT and most presumably 
non-distillable.}
\label{distill}
\end{figure}

\appendix

\section{Sufficient criterion for inseparability}

Our sufficient criterion of inseparability (i.e. a necessary criterion for separability) is based on the following
Lemma. Let us assume the product vectors $|e,f\rangle$ to be normalized.

\newtheorem{lemm1}{Lemma}
\begin{lemm1}
If max$\{{\rm Tr}(\rho^2), {\rm Tr}((\rho^{T_B})^2), {\rm Tr}(\rho \rho^{T_B})\} > \mbox{max}_{|e,f\rangle} 
\langle e,f|\rho|e,f\rangle$, then $\rho$ is inseparable.
\end{lemm1}
Let us prove that this statement is true in the case where $\{{\rm Tr}(\rho^2)=$
 max$\{{\rm Tr}(\rho^2), {\rm Tr}((\rho^{T_B})^2), {\rm Tr}(\rho \rho^{T_B})\}$. All the other
cases can be proved in the same way which implies that we can take the 
maximum of those three values.

\noindent{\em Proof:} We define $r=\mbox{max}_{|e,f\rangle}\langle
 e,f|\rho|e,f\rangle$ and the witness, $X=1-\frac{1}{r}\rho$.
 Assuming that ${\rm Tr}(\rho)=1$ we have that ${\rm Tr}(X\rho)<0$. 
It remains to prove that ${\rm Tr}(X\tilde{\rho})>0$ for all $\tilde{\rho}$
separable. We write $\tilde{\rho}= \sum_i \lambda_i |e_i,f_i\rangle
\langle e_i,f_i|$, and observe that: $r{\rm Tr}(X\tilde \rho)={\rm Tr}[(\mbox{max}_{|e,f\rangle}
\langle e,f|\rho|e,f\rangle-\rho)\tilde \rho]=\sum_i \lambda_i 
(\mbox{max}_{|e,f\rangle}\langle e,f|\rho|e,f\rangle-\langle e_i,f_i|\rho|e_i,f_i\rangle)\geq 0$.

Unfortunately this criterion does not work for the Horodecki PPT states in $2\times4$ space \cite{pawel}. It does, however, detect
the entanglement of the PPT states constructed from UPB's\cite{UPB}. In such case $\rho=P/K$, where $P$ is a projector 
onto a space
that does not contain any product vectors, $K=r(P)$, and $P=P^{T_B}$. Our criterion gives 
${\rm Tr}(\rho^2)={\rm Tr}((\rho^{T_B})^2)= {\rm Tr}(\rho \rho^{T_B})=1/K$, while $\langle e,f|\rho|e,f\rangle=\langle
e,f|P|e,f\rangle/K<1/K$. 
 
\section{Separability check for binary mixtures of product states}

If a separable matrix is a mixture of two product states
(here we call such matrix a binary mixture), then it is
relatively easy to check separability. Assume $\rho$ of the following form:
\be
\rho=\sum_{i=1}^{K}p_i \rho_i^{A}\otimes \rho_i^B,
\label{werner1}
\ee
where all $p_i>0$, and $\sum_ip_i=1$. Let $\mu$ and $\nu$ be  density
matrices acting in Alice and Bob's space respectively. Let us define the
matrix function
\be
M(\mu,\nu)=\sum_{i=1}^{K}p_i (\rho_i^{A}-\mu)\otimes (\rho_i^B-\nu).
\label{M}
\ee
Interestingly, $M$ can be calculated without the explicite use of 
the representation (\ref{werner1}),
\be
M(\mu,\nu)=\rho-\mu\otimes\rho_B-\rho_A\otimes\nu+\mu\otimes\nu,
\label{MM}
\ee
where the reduced density matrices are $\rho_{A,B}={\rm Tr}_{B,A}\rho$.

For $K=2$ we have an obvious Lemma:

\begin{lemm1}
If $\rho$ is a mixture of two product states
(with  $p_1=p, p_2=1-p$) then the equation $M(\mu,\nu)=0$ has at least
two solutions $\mu=\rho^A_1,\ \nu=\rho^B_2$, and $\mu=\rho^A_2,\
\nu=\rho^B_1$.
\end{lemm1}
The ``opposite" implication is also true.

\begin{lemm1}
 If the equation $M(\mu,\nu)=0$ has 
solutions such that $\mu\ge 0$, $\mu=\mu^{\dag}$, ${\rm Tr}\mu=1$,
$\nu\ge 0$, $\nu=\nu^{\dag}$, ${\rm Tr}\nu=1$, then $\rho$ is either a
separable binary mixture, or a nonseparable binary pseudomixture.
\end{lemm1}
The proof follows directly from Eq. (\ref{MM}). $\rho$ can be written
as
\be
\rho=\mu\otimes(\rho_B-(1-p)\nu)+(\rho_A-p\mu)\otimes\nu,
\label{MMM}
\ee
with some $0\le p\le 1$. We immediately see that $\rho$ is separable if
there exists $p\in[0,1]$ such that $\rho_B-(1-p)\nu\ge 0$ and
$\rho_A-p\mu\ge 0$. This implies that the ranges $R(\mu)$ and $R(\nu)$
must be included in the ranges of $\rho_A$, and $\rho_B$, respectively;
at the same time $p$ must fulfill the conditions
$p||\mu(\rho_A)^{-1}||\le 1$, and $(1-p)||\mu(\rho_B)^{-1}||\le 1$,
where $||.||$ denotes the operator norm (for details see \cite{2xn}).

Checking if the equation $M(\mu,\nu) =0$ has solutions is very easy. We
can use a product basis $\{O_i^A\otimes O^B_j\}_{i=1,\ldots,M^2;
j=1,\ldots,N^2}$ in the operator space. Such basis can be chosen to be
orthonormal and hermitian with respect to the trace scalar product. 
The equation $M(\mu,\nu)=0$ projected onto the $ij$--th element of the
basis reads:
\be
\rho_{ij}-\mu_i\rho_{Bj}-\rho_{Ai}\nu_j+\mu_i\nu_j=0.
\ee
where $\rho_{ij}={\rm Tr} \rho O_i^A\otimes O_j^B$, $\rho_{Ai}={\rm Tr}\rho_AO_i^A$,  $\rho_{Bj}={\rm Tr}\rho_B O_j^A$, etc...We have thus $N^2M^2$ such equations for $M^2+N^2$ real coefficients
$\mu_i$, $\nu_j$. The equations have a very simple structure and
therefore it is easy to check:
a) if they have a solution; b) if the resulting $\mu$ and $\nu$ are
positive definite; c) if there exists $p$ such that both terms on the RHS of
Eq. (\ref{MMM}) are positive definite. 

The above formulated separability check can be easily generalized to
systems of $R$ parties and separable mixtures of $R$ product states
for $R>2$.

\section{Finding the {\em local} minimum for projecting onto the
2-dimensional subspace}
In this appendix we will find the states that lead to
a local minimum 
for projection  onto a two-dimensional subspace
as in equation (\ref{sub})
in the case of 2 copies (in dimension $3\times 3$), 
for the critical value of the parameter $\beta=3/2$. 
\par
We will proceed as follows: our problem  will be formulated
in terms of a function $f(\lambda, \psi)$ that has to be minimized
with respect to $\psi$. We will find a family of states $\psi$ 
for which $f$ is shown to reach a local minimum, for a range of parameters
$\lambda$.
\par
We introduce a parameter $\lambda$ and  will study 
 the following
function:
\bea
f(\lambda, \psi) &=& \frac{1}{M^2}
                \bra{\psi} (1+\abs{1-2\lambda})\eins\otimes\eins+
                \frac{9}{4}P\otimes P \nonumber\\&-&3 
                (\lambda \cdot \eins\otimes P
                +(1-\lambda)P\otimes\eins)\ket{\psi}  \nonumber\\
                &=&  \bra{\psi}\rho_2(\lambda)\ket{\psi}
                \ ,
\label{function}
\eea
where the last line defines $\rho_2(\lambda)$, and $P$ is the projector onto a maximally entangled state.
Here $\lambda$ is a given fixed parameter with 
$0<\lambda<1$. The case $\lambda=\half$ 
corresponds to $\rho_2$ being the partial transpose of two copies of $\eins-3P/2$. In the notation of 
Section \ref{dist} this corresponds
to the value of $\beta=3/2$. We are looking for
the minimum of $f$ with respect to $\ket{\psi}$. This state lives
in the two-dimensional subspace and can be written in
 the Schmidt decomposition (cf. \cite{Peres})
\be
\ket{\psi} = a \ket{e_1}_A\ket{f_1}_B +b\ket{e_2}_A\ket{f_2}_B\ ,
\label{state}
\ee
where the states $\ket{e_i},\ket{f_i}$ are normalised and
$\bra{e_1^*}e_2\rangle = 0 = \bra{f_1^*}f_2\rangle$ and 
$\abs{a}^2+\abs{b}^2=1$. For clarity we kept the indices $A$ and $B$.
\par
Let us rewrite the terms in equation (\ref{function}), sorting
them not pairwise, but with respect to Alice and Bob:
\bea
\eins\otimes\eins &=& \sum_{i,j,r,s=1}^3 \ket{ir}_A
                      \ket{js}_B\bra{ir}_A\bra{js}_B, \nonumber \\
P\otimes P &=& \frac{1}{9}\sum_{i,j,r,s=1}^3 
                      \ket{ir}_A\ket{ir}_B\bra{js}_A\bra{js}_B, \nonumber \\
P\otimes\eins &=& \frac{1}{3}\sum_{i,j,r,s=1}^3
                      \ket{ir}_A\ket{is}_B\bra{jr}_A\bra{js}_B, \nonumber \\
\eins\otimes P &=& \frac{1}{3} \sum_{i,j,r,s=1}^3 
                      \ket{ir}_A\ket{jr}_B\bra{is}_A\bra{js}_B \ .
\label{albo}
\eea
This notation fixes the basis in which we will also write $\ket{\psi}$.
Indices $i,j$ are used for the first pair and $r,s$ for the second pair.
\par
The minimum of $f(\lambda=0,\psi)$ is found by requiring the two conditions
\begin{itemize}
\item[a)] $\bra{\psi}P\otimes P\ket{\psi} =0 $,
\item[b)]$\bra{\psi}P\otimes\eins\ket{\psi}$  maximal .
\end{itemize}
According to equations (\ref{albo})
we reach a) only if the entries of either the first or the second pair
are orthogonal to each other. We can reach b) if the
entries in the first bits of Alice and Bob
 are identical in both terms of the 
Schmidt decomposition of $\ket{\psi}$, and if their second bits are in a 
product state.
The coefficient  in equation (\ref{state})
is easily found to be $\abs{\alpha}=1/\sqrt{2}$
for maximisation of b). 
\par
We can fulfill both conditions a) and b) with
the family of states that minimizes $f$ for $\lambda=0$, denoted
by $\ket{\psi^\star}$:
\be
\ket{\psi^\star}=\shalf(\ket{ir}_A\ket{is}_B+e^{i\varphi}\ket{jr}_A\ket{js}_B),
 \label{family}
\ee
with $\scal{i}{j} =0=\scal{r}{s}$.
Therefore for $\lambda=0$ we have found the {\em global} minimum of $f$
to be 
\be
\min_{\psi}f(\lambda =0)=0 \ .
\ee
Using the explicit structure of the states $\ket{\psi^\star}$
we find that 
\bea
\bra{\psi^\star}\eins\otimes\eins\ket{\psi^\star}&=&1 \ , \nonumber \\
\bra{\psi^\star}\eins\otimes P\ket{\psi^\star}&=&0 \ , \nonumber \\
\bra{\psi^\star}P\otimes\eins\ket{\psi^\star}&=&2/3 \ .
\eea
When varying the parameter $\lambda$ we therefore find
\be
\bra{\psi^\star}\rho_2(0\leq\lambda\leq\half)\ket{\psi^\star}=0 \ .
\ee
Similarly,
for $\half\leq\lambda\leq 1$ the same line of argument holds when
interchanging the role of first and second bits, leading to a 
different minimizing family $\ket{\psi^\bullet}$. At $\lambda=\half$,
the point which is symmetric with respect to interchanging the
two pairs, both
families lead to the expectation value zero.
\par 
In the following we will show that $\ket{\psi^\star}$ corresponds to
  a local  minimum for $0<\lambda\leq\half$.
\par
First, we 
show that the states given in equation
(\ref{family}) form a compact set, by
describing how to move through the whole family  in infinitesimal steps:
 Looking at the first pair, we can either make the
following change:
\be
\ket{i}\rightarrow x_i \ket{i} +x_k \ket{k} \ \ \rm{with} \
                 \scal{k}{i}=0=\scal{k}{j} \ ,
\ee
or we can move to
\be
\ket{j}\rightarrow  x_j\ket{j} +x_l \ket{l} \ \ \rm{with} \
                 \scal{l}{j}=0=\scal{l}{i} \ ,
\ee
or we can change both $\ket{i}$ and $\ket{j}$, keeping $\scal{i}{j}=0$.
Regarding the second pair, we can change
\be
\ket{r}\rightarrow x_r \ket{r} +x_p \ket{p} \ \ \rm{with} \
                 \scal{p}{r}=0=\scal{p}{s} \ ,
\ee
or we can move to 
\be
\ket{s}\rightarrow x_s \ket{s} +x_t \ket{t} \ \ \rm{with} \
                 \scal{t}{r}=0=\scal{t}{s} \ .
\ee
In this way we can move within the family in infinitesimal steps,
and there are no isolated points.
\par
Let us now move outside of our family by an infinitesimal amount.
We will write down the most general path leading away from the family
and then show that first order terms of the expectation value vanish,
i.e. we have an extremum, and that the functional determinant of second
order terms is positive, i.e. we have  a local minimum.
\par
The most general infinitesimal step away from our family is given by
\bea
&&\ket{\psi^\star+\delta}=
\shalf(\sqrt{1+\delta_0}\ket{ir}_A\ket{(i+\delta_1 k)(s+\delta_2 r)}_B
   + \nonumber \\
  && \sqrt{1-\delta_0}\, e^{i\varphi}\ket{(j+\delta_3 l)(r+\delta_4 t)}_A
   \ket{(j+\delta_5 m)(s+\delta_6 r)}_B)\nonumber \\
  && \ \   \rm{with} \ \ \ \scal{k}{i} =0=\scal{l}{j}, \nonumber \\
 && \ \ \ \ \ \ \ \ \ \ \ \, \scal{l}{i}=0=\scal{t}{r} ,  \nonumber \\
   && \ \ \ \ \ \ \ \ \ \,\, \scal{m}{j}=0=\scal{i+\delta_1 k}{j+\delta_5 m} \ ,
\eea
so that the Schmidt terms are still orthogonal, and for each of the seven
$\delta_i\neq 0$ we leave the family. Note that we can always keep one
state (in this case $\ket{ir}_A$) constant by using bilateral rotations.
\par
We now expand the expectation value
\bea
\langle \varrho_2\rangle &=& \frac{1}{M^2}\bra{\psi^\star+\delta}
           \varrho_2 (0<\lambda\leq\half) 
        \ket{\psi^\star+\delta}\ \  
\eea
in powers of $\delta_i$, and find that all terms linear in $\delta_i$
are indeed vanishing.
\par
The second order terms can be written down explicitly. The diagonal ones are:
\bea
\calo(\delta_0^2): \ \ & & \half \delta_0^2(1-\lambda) > 0 \nonumber \\
\calo(\delta_1^2): \ \ & & \half \delta_1^2(1+\abs{1-2\lambda}) > 0 \nonumber \\
\calo(\delta_2^2): \ \ & & \half \delta_2^2(1+\abs{1-2\lambda}
                     +\frac{1}{4}-1) > 0 \nonumber \\
\calo(\delta_3^2): \ \ & & \half \delta_3^2(1+\abs{1-2\lambda}) > 0 \nonumber \\
\calo(\delta_4^2): \ \ & & \half \delta_4^2(1+\abs{1-2\lambda}
                     +\frac{1}{4}\delta_{ts}-(\lambda \delta_{ts}+1-\lambda)) 
                           > 0 \nonumber \\
\calo(\delta_5^2): \ \ & & \half \delta_5^2(1+\abs{1-2\lambda}) > 0 \nonumber \\
\calo(\delta_6^2): \ \ & & \half \delta_6^2(1+\abs{1-2\lambda}
                     +\frac{1}{4}-1) > 0 \ .
\label{diag}
\eea
Here $\delta_{ts}$ is the Kronecker symbol.
\par
Nearly all off-diagonal  terms of second order 
vanish, the only non-zero one   is
 [note that $\calo(\delta_i\delta_j)=\calo(\delta_j\delta_i)$]:
\be
\calo(\delta_2\delta_6): \ \  
          \half \delta_2\delta_6(\lambda-\frac{3}{4})  \ .
\label{offdiag}
\ee
This term is negative
for the range $0<\lambda\leq\half$; the corresponding 2x2 determinant, however, is 
${\rm det}(2,6)=3\lambda^2-\frac{7}{2}\lambda+1 \geq 0$ for $0<\lambda\le 1/2$. Thus we have found
the second derivative to be positive, and therefore our family
$\ket{\psi^\star}$
corresponds to a local minimum for $0<\lambda\leq\half$. In particular that is the case for 
$\lambda=\half$, i.e. $\beta=3/2$ in the
notation of Section \ref{dist}.  
It is easy to see that this must also be the case for all $1\ge \beta\le3/2$.
\par
For a complete proof that two copies are not distillable, however,
it remains to be shown that this minimum is
 a {\em global} minimum for  $\beta=3/2$.


\end{document}